\documentclass[pra,aps,twocolumn,showpacs,superscriptaddress,10pt,floatfix]{revtex4-1}

\usepackage{graphicx}
\usepackage{amsmath,amsfonts,amssymb,color,mathptmx}
\usepackage{bm}
\usepackage[utf8]{inputenc}
\usepackage[colorlinks=true,linkcolor=blue,citecolor=blue,urlcolor=black]{hyperref}

\newcommand{\sinc}{\ensuremath{\mbox{\hspace{1.3pt}sinc}}}

\newcommand{\Si}{\ensuremath{\mbox{\hspace{1.3pt}Si}}}

\begin{document}


\title{Non-Gaussian state generation certified using the EPR-steering inequality}

\author{E. S. G\'{o}mez}
\email{estesepulveda@udec.cl}
\author{G. Ca\~nas}
\author{E. Acu\~na}
\affiliation{Center for Optics and Photonics, MSI-Nucleus on Advanced Optics, Departamento de F\'{\i}sica, Universidad de Concepci\'{o}n, 160-C, Concepci\'{o}n, Chile}
\author{W. A. T. Nogueira}
\affiliation{Departamento de F\'isica, ICE, Universidade Federal de Juiz de Fora, Juiz de Fora, CEP 36036-330, Brazil}
\affiliation{Universidade Federal de Minas Gerais,
 Caixa Postal 702, 30123-970, Belo Horizonte, MG, Brazil}
\author{G. Lima}
\affiliation{Center for Optics and Photonics, MSI-Nucleus on Advanced Optics, Departamento de F\'{\i}sica, Universidad de Concepci\'{o}n, 160-C, Concepci\'{o}n, Chile}
\date{\today}



\begin{abstract}
Due to practical reasons, experimental and theoretical continuous-variable (CV) quantum information (QI) has been heavily based on Gaussian states. Nevertheless, many CV-QI protocols require the use of non-Gaussian states and operations. Here, we show that the Einstein-Podolsky-Rosen steering inequality can be used to obtain a practical witness for the generation of pure bipartite non-Gaussian states. While the scenario require pure states, we show its broad relevance by reporting the experimental observation of the non-Gaussianity of the CV two-photon state generated in the process of spontaneous parametric down-conversion (SPDC). The observed non-Gaussianity is due only to the intrinsic phase-matching conditions of SPDC.
\end{abstract}

\maketitle

\section{Introduction}
Continuous-variable (CV) quantum information (QI) is a research field that has been increasingly growing in the past few years \cite{Braunstein}. The need to cover larger Hilbert spaces is motivated by QI protocols, such as quantum key distribution (QKD), that has advantages when implemented in higher dimensions \cite{Gisin1,Lima13}. The extension of many QI protocols, first proposed considering discrete quantum systems, to the realm of CV systems has been heavily based on Gaussian states \cite{Eisert1,Cirac,Acin}. This is due to the fact that their covariance matrices are fully determined by the first and second order moments \cite{Paris1,Weedbrock}, and also because of the practicality in the creation, manipulation and detection of Gaussian states \cite{Vaidman,CV-QKD}.

Nevertheless, several recent works have shown the relevance of non-Gaussian states and operations \cite{SteveNG}. For instance, they are required for quantum computation with cluster states \cite{Lund08_1}, entanglement distillation \cite{Eisert02_1,Fiurasek02_1}, quantum error correction \cite{Fiurasek09_1} and loophole-free Bell tests \cite{Banaszek99_1,Carmichael04_1}. Besides, they provide advantages to the quantum teleportation, quantum cloning and state estimation tasks \cite{Bonifacio03_1,Cerf05_1,Genoni09_1,Adesso09_1}.

In this work, we show that the Einstein-Podolsky-Rosen (EPR) steering inequality \cite{EPR,Schro,Reid,Steer} can be used to obtain a witness for the non-Gaussianity of pure bipartite CV quantum states. While the scenario require pure states, we show its broad relevance by reporting the observation of the non-Gaussianity of the CV two-photon state generated in the process of spontaneous parametric down-conversion (SPDC)\cite{Mandel,Monken1,Saleh1,PR}. SPDC is up to date the most used source for experimental investigations in the field of quantum information, and our work highlights the simplicity of using this source for new applications in CV-QI. The generated down-converted photons are correlated in their transverse momenta and can be used to test the EPR-paradox \cite{BoydEPR}. The observed non-Gaussianity is due only to the intrinsic phase-matching conditions of the SPDC process \cite{TboNG,ExterNF2}, thus, highlighting the simplicity of using SPDC sources for new applications of CV-QI.


\section{The EPR-steering inequality with bipartite Gaussian states}\label{teoria}
Consider a bipartite system described by a pure CV-state $\rho_{12}=|\psi\rangle\langle\psi|$, and a generic pair of complementary noncommuting observables $\hat{u}_i$ and $\hat{v}_i$, with $i=1,2$ used to denote the operation at each subsystem. The spectral decomposition of $\hat{u}_i$ and $\hat{v}_i$ is a infinite set of continuous variables. In Ref.~\cite{Reid} it was introduced the EPR-steering criterion as $\Delta_{inf}^2(\hat{u}_2)\Delta_{inf}^2(\hat{v}_2)\geq C$, where $C$ is is a value that depends on the chosen observables. The violation of this inequality implies the implementation of a EPR-paradox \cite{BoydEPR,Steve}. The \emph{inferred} variances are given by
\begin{equation}\label{eq:def_varinf}
\Delta_{inf}^2(\hat{u}_2)=\int du_1\, P(u_1)\Delta^2(u_2|u_1),
\end{equation}
where $\Delta^2(u_2|u_1)$ is the variance of the conditional probability distribution $P(u_2|u_1)$, and $P(u_1)$ is the marginal probability distribution of one party's outcomes.

Now, let us consider a general continuous variable two mode pure state described by a Gaussian amplitude $\mathcal{A}(\mathbf{q}_1,\mathbf{q}_2)$, where the vectors $\mathbf{q}_1$ and $\mathbf{q}_{2}$ are the outputs of the observable $\hat{q}$ on each party
\begin{equation}\label{gstate}
\mathcal{A}_G(\mathbf{q}_1,\mathbf{q}_2)\propto\exp\left(-\frac{|\mathbf{q}_1+\mathbf{q}_2|^2}{4\sigma_+^2}\right)\exp\left(-\frac{|\mathbf{q}_1-\mathbf{q}_2|^2}{4\sigma_-^2}\right),
\end{equation}
with $\sigma_+$ and $\sigma_-$ being the widths of the corresponding Gaussian functions. Several CV physical systems can be modelled in this way, for example: an atomic ensemble interacting with an electromagnetic field, two entangled photons sent through Gaussian channels, photoionization of atoms and photodissociation of molecules \cite{Fedorov2004}, spontaneous emission of a photon by an atom \cite{Eberly2002,Eberly2003,Fedorov2005}, and multiphoton pair production \cite{Fedorov2006,Eisert}. Since Gaussian amplitudes have the same functional form in both transverse directions we can, without loss of generality, work in one dimension and consider only its scalar form.

\begin{figure}[t]
\centering
\includegraphics[angle=0,width=0.48\textwidth]{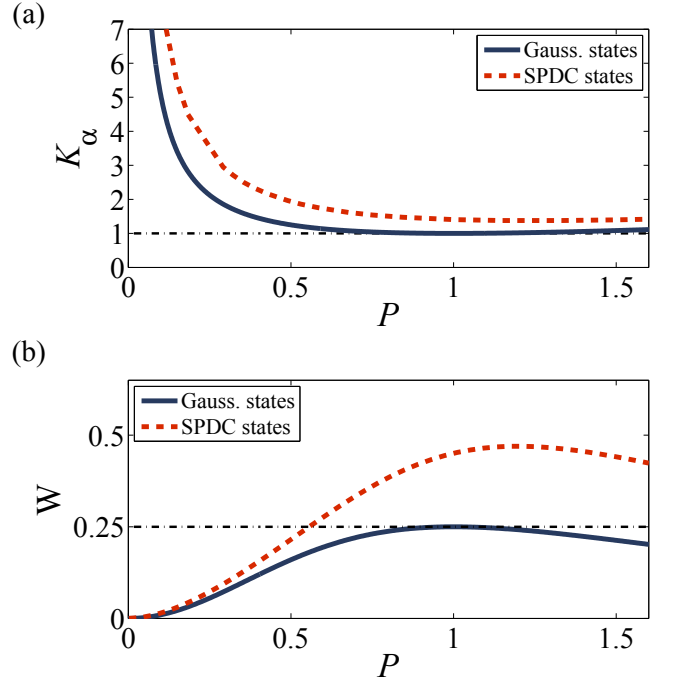}
\caption{(Color Online) (a) Shows the Schmidt number for bipartite Gaussian states and for the CV spatial state of SPDC [Eq.~(\ref{SPDC-mom})], while varying $P$ and considering one transverse direction of $\mathbf{q}_j$. (b) The values of $W$ plotted in terms of $P$.
\label{Fig1}}
\end{figure}

From Eq.~(\ref{eq:def_varinf}) one can obtain the limit of $C$ for Eq.~(\ref{gstate}) \cite{Eisert}. Note that $\Delta_{inf,G}^2(\hat{q}_2)=\Delta^{2}(q_{2}|q_{1})=\sigma_{+}^{2}\sigma_{-}^{2}/(\sigma_{+}^{2}+\sigma_{-}^{2})$. One can also find the inferred variance of the complementary observable $\hat{x}_{2}$, and it is given by $\Delta_{inf,G}^2(\hat{x}_2)=\Delta^{2}(x_{2}|x_{1})=1/(\sigma_{+}^{2}+\sigma_{-}^{2})$. Therefore, the EPR-steering inequality reads
\begin{equation}\label{EPRgauss}
\Delta^{2}(q_{2}|q_{1})\Delta^{2}(x_{2}|x_{1})\!=\!\frac{\sigma_{+}^{2}\sigma_{-}^{2}}{\left(\sigma_{+}^{2}+\sigma_{-}^{2}\right)^2}\geq\frac{1}{4}.
\end{equation}
Let us define one parameter $P\equiv\sigma_{+}/\sigma_{-}$ for simplicity. The Schmidt Number $K_{G}$ for the general Gaussian state of Eq.~(\ref{gstate}) is given by \cite{Eberly}
\begin{equation}\label{schmidtgauss}
K_{G}=\frac{1}{4}\left(\frac{1}{P}+P\right)^2.
\end{equation}
Thus, the EPR-steering inequality can be written as
\begin{equation}\label{EPR-gauss1}
\Delta^{2}(q_{2}|q_{1})\Delta^{2}(x_{2}|x_{1})=\frac{1}{4K_{G}}\geq\frac{1}{4}.
\end{equation}
Note that due to the symmetry of Eq.~(\ref{gstate}) one has that $K_G=K_{G\alpha}\times K_{G\beta}$, where $K_{Gj}$ represents the Schmidt number in the transverse direction $j=\alpha,\beta$ \cite{Kfact}. Thus, one may further simplify the EPR-steering inequality for Gaussian states to
\begin{equation}\label{EPR-gauss2}
W\equiv\Delta^{2}(q_{2}|q_{1})\Delta^{2}(x_{2}|x_{1})=\frac{1}{4K_{G\alpha}^2}\geq\frac{1}{4}.
\end{equation}
Thus, one may test it by performing the measurements in only one transverse direction. Moreover, it does not depend of the values chosen for $q_{1}$ and $x_{1}$. When $P=1$, $K_{G}=K_{G_\alpha}=K_{G_\beta}=1$ and the Gaussian state is a product state. In Fig.~\ref{Fig1}(a) [Fig.~\ref{Fig1}(b)] we show $K_{G_\alpha}$ ($W$) with a solid blue line, while varying $P$.


\section{A witness for the non-Gaussianity of CV quantum states} \label{SPDC}
From the results obtained above it is possible to envisage a simple and practical way to determine if a certain bipartite pure state is Gaussian or not. Note from Eq. (\ref{EPR-gauss2}) and Fig.~\ref{Fig1}(b) that pure bipartite Gaussian states will always violate the EPR-steering inequality. The reason is that they are in general entangled states [See Fig.~\ref{Fig1}(a)]. The only exception is the point marked with the horizontal dashed line in Fig.~\ref{Fig1}(b), which represents the point where the Gaussian state is a product state. In this case, we have that $W=\frac{1}{4K_{G}}=0.25$, which corresponds to the upper quantum bound for pure bipartite Gaussian states. Thus, the observation of a value greater than 0.25 for $W$ with pure entangled states can only be achieved while considering non-Gaussian states.


\begin{figure*}[t]
\centering
\includegraphics[angle=0,width=1\textwidth]{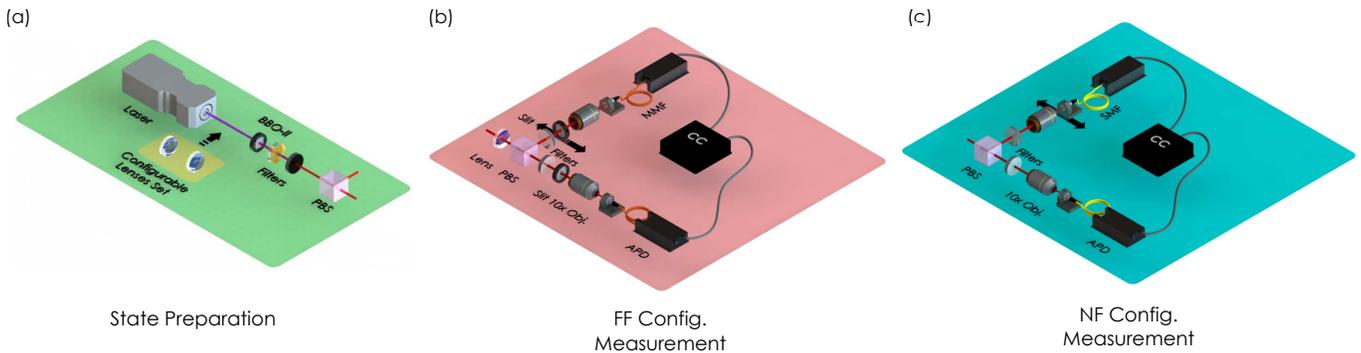}
\caption{(Color Online) Experimental setup. (a) State preparation stage. (b) Setup configuration for measuring $\Delta^{2}(q_{2}|q_{1})$. (c) Setup configuration for measuring $\Delta^{2}(x_{2}|x_{1})$. See the main text for details. \label{Fig2}}
\end{figure*}


While the scenario require pure states, now we show its broad relevance by reporting the observation of the non-Gaussianity of the CV spatial two-photon state generated in the process of SPDC \cite{Mandel,Monken1,Saleh1,PR}. When perfect colinear phase-matching is considered and neglecting effects of anysotropy, we can write the spatial two-photon state as \cite{Monken1,Saleh1}
\begin{equation}\label{SPDC-mom}
|\psi\rangle_{12}\propto\iint d\mathbf{q}_1 d\mathbf{q}_2\,\tilde{E}_{p}(\mathbf{q}_1+\mathbf{q}_2)\tilde{G}(\mathbf{q}_1-\mathbf{q}_2)|1\mathbf{q}_1\rangle|1\mathbf{q}_2\rangle,
\end{equation}
where $|1\mathbf{q}_i\rangle$ represents one photon in mode $i$ ($i=1,2$) usually called signal or idler, and with the transverse momentum $\mathbf{q}$. $\tilde{E}_{p}(\mathbf{q})$ is the angular spectrum of the pump beam. Usual experimental configurations adopt a Gaussian pump beam and in this case $\tilde{E}_{p}(\mathbf{q})\propto\exp\left[-c^2 |\mathbf{q}|^2/4\right]$. $c$ represents the beam radius at the crystal plane. $\tilde{G}(\mathbf{q})=\sinc\left(b|\mathbf{q}|^2\right)$ defines the phase-matching conditions of the SPDC process, with $\sinc(\xi)\equiv\frac{\sin(\xi)}{\xi}$. $b$ is defined by $b\equiv\frac{L}{8k}$, where $L$ is the crystal length, and $k$ the wavenumber of the down-converted photons. In terms of these definitions, $P$ reads $\frac{1}{c}\sqrt{\frac{L}{2k}}$. This state can be rewritten in the complementary transverse position representation as \cite{ExterNF2,Monken2}
\begin{equation}\label{SPDC-pos}
|\psi\rangle_{12}\propto\iint d\boldsymbol{x}_1 d\boldsymbol{x}_2\,E_{p}\left(\frac{\boldsymbol{x}_1+\boldsymbol{x}_2}{2}\right)G\left(\frac{\boldsymbol{x}_1-\boldsymbol{x}_2}{2}\right)|1\boldsymbol{x}_1\rangle|1\boldsymbol{x}_2\rangle,
\end{equation}
where the functions $E_{p}(\boldsymbol{x})$ and $G(\boldsymbol{x})$ are the Fourier transform of $\tilde{E}_{p}(\mathbf{q})$ and $\tilde{G}(\mathbf{q})$. Thus, $E_{p}(\boldsymbol{x})\propto \exp\left[-|\boldsymbol{x}|^2/c^2\right]$ and $G(\boldsymbol{x})\propto 1-\frac{2}{\pi}\Si\left(\frac{1}{4b}|\boldsymbol{x}|^2\right)\equiv\mathrm{sint}\left(\frac{1}{4b}|\boldsymbol{x}|^2\right)$, where $\Si(x)\equiv\int_0^x dt\, \sinc(t)$. Here, the transverse position $\hat{x}_j$ and transverse momentum $\hat{q}_j$ are the complementary observables for the EPR-steering inequality test \cite{BoydEPR}. Clearly, the CV spatial two-photon state of SPDC is a non-Gaussian state \cite{TboNG}.

An experimental observation of the effects that arise form the phase-matching conditions, namely, the non-Gaussianity of $|\psi\rangle_{12}$, was reported in Ref. \cite{ExterNF2}. They demonstrated how the spatial correlations in the near field plane of a non-linear crystal changes when the phase matching conditions varies. Now, we demonstrate experimentally how $W$ can be used to detect the non-Gaussianity of this state. The demonstration is based on the fact that the spatial state of SPDC process is pure and entangled \cite{Eberly}, even considering a post-selected one transverse direction \cite{TboNG,PR}. This means that for any value of $P$, the Schmidt number $K_{S\alpha}$ is always greater than 1. In Fig.~\ref{Fig1}(a) [Fig.~\ref{Fig1}(b)] we show $K_{S_\alpha}$ ($W$) with a dashed red line, while varying $P$ for the state $|\psi\rangle_{12}$. One can see that for some values of $P$ ($0.56\leq P\leq 2.58$) the values of $W$ are greater than $\frac{1}{4}$, thus, witnessing the non-Gaussianity of this state.

\section{Experiment}\label{resultados}

The experimental setup is illustrated in Fig.~\ref{Fig2}. We used a solid-state laser source, at $355\,nm$, to pump a $\beta$-barium-borate type-II (BBO-II) non-linear crystal ($L= 1.8\,cm$) for the generation of the down-converted photons. Initially, the Gaussian pump beam had a waist of $c=200\,\mu m$ at the crystal plane. However, to experimentally observe the dependence of $W$ with $P$ [See Fig.~\ref{Fig1}(b)], we generated five more different states by changing the waist of the beam at the crystal plane. This has been done by using a configurable set of doublet achromatic lenses placed before the non-linear crystal [See Fig.~\ref{Fig2}(a)]. The corresponding values of $P$ for the six generated states are shown in Tab.~\ref{tabla1:estados}.

\begin{table}[h]
\caption{Corresponding values of $P$ for the generated states.\label{tabla1:estados}}
\begin{ruledtabular}
\begin{tabular}{ccc}
State & $c\,[\mu m]$ & $P$\\ \hline
1 & 200 & 0.1595 \\
2 & 100 & 0.3189 \\
3 & 70 & 0.4556 \\
4 & 45 & 0.7087\\
5 & 40 & 0.7973 \\
6 & 35 & 0.9112 \\
\end{tabular}
\end{ruledtabular}
\end{table}

To guarantee the purity of the measured states, we used spatial and spectral filters in each measurement apparatus. For instance, interference filters were used to select degenerated down-converted photons at $710\,nm$ with $5\,nm$ of bandwidth. This fact implies no entanglement between the frequencies and the transverse spatial coordinates, and then the reduced spatial state must be pure. Besides, the usage of spatial filters is explained afterwards. 

Furthermore, a polarizer beam splitter (PBS) separates the signal and idler photon modes. For measuring  $\Delta^{2}(q_{2}|q_{1})$ and $\Delta^{2}(x_{2}|x_{1})$, we performed conditional coincidence measurements between the idler and signal photons in two different transverse planes (See Fig.~\ref{Fig2}): the far- and near-field planes of the non-linear crystal, respectively \cite{BoydEPR}.

\subsection{Measurement of $\Delta^2(q_2|q_1)$}
In order to compute $\Delta^2(q_2|q_1)$, one shall perform conditional coincidence measurements at the far-field plane of the non-linear crystal. The reason is very simple: at this plane, the coincidence rate $C_q(x_1,x_2)$ is given by
\begin{equation}\label{Eq_coincFF}
C_{q}(x_1,x_2)\propto\left|\tilde{E}_{p}\left[\frac{k}{f_{q}}(x_1+x_2)\right]\tilde{G}\left[\frac{k}{f_{q}}(x_1-x_2)\right]\right|^2,
\end{equation} and, since $q_j=k x_j/f_{q}$, it maps the square of the probability amplitude of Eq.~(\ref{SPDC-mom}), i.e., the momentum correlation of the photons generated in the SPDC process.

Figure~\ref{Fig2}(b) shows our setup configuration. One lens $L_{q}$ with focal distance $f_{q}=15\,cm$ was placed before the PBS to create the far-field plane for both signal and idler beams. At these planes, vertical slits were placed for post-selecting the desirable state \cite{LeoGen,LVNRS09} and performing the conditional coincidence measurements \cite{SteveNG,BoydEPR,Yuan}. The width of each slit is $50\,\mu m$.  The effect of using a non point-like detector is that the transmittance function of the slit may broad the far-field distribution to be measured. However, in the case of 50 $\mu m$ slits and the experimental configuration adopted, this effect is negligible \cite{Kfact}. This can be easily checked through the calculation of the convolution between the transmittance function of the slit and the predicted far-field distribution. 

After the transmission through each slit, the down-converted photons were collected with a $10$x objective lens and multi-mode fibers. The fibers were connected to single-photon counting modules, and then a coincidence circuit (with $4\,ns$ of coincidence window) recorded the data. To perform the coincidence conditional measurements of $\Delta^2(q_2|q_1)$, we scanned in the horizontal direction one slit (of mode 2) while the other one was fixed at the center ($q_1=0$).

\begin{figure}[t]
\centering
\includegraphics[angle=270,width=0.45\textwidth]{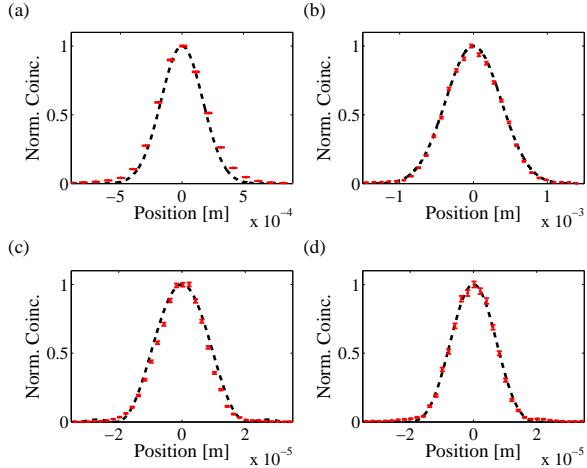}
\caption{(Color Online) Experimental measurement of the far- and near-field conditional distributions. In (a) [(c)] we show the results for the second state at the far-field (near-field) plane. In (b) [(d)] we show the results for the sixth state at the far-field (near-field) plane. The (red) dots represent the experimental data and the dotted (black) lines are the theoretical predictions from Eq.~(\ref{Eq_coincFF}) and Eq.~(\ref{Eq_coincNF}). \label{Fig3}}
\end{figure}

To give an example of the results obtained while scanning the slit at mode-2, we show in Fig.~\ref{Fig3}(a) [Fig.~\ref{Fig3}(b)] the far-field conditional distribution measured for the second (sixth) state generated. The experimental results are represented by red points (error bars lie inside the points due to the observed high rate of coincidence counts) and the black dotted-line is the theoretical curve for these distributions arising from Eq.~(\ref{Eq_coincFF}).


\subsection{Measurement of $\Delta^2(x_2|x_1)$}
To measure $\Delta^2(x_2|x_1)$ it is now necessary to measure at the near-field plane of the non-linear crystal. Again, the reason is very simple: at this plane, the coincidence distribution is proportional to the square of the amplitude of Eq. (\ref{SPDC-pos}), that is
\begin{equation}\label{Eq_coincNF}
C_{x}(x_1,x_2)\propto\left|E_{p}\left(\frac{x_1+x_2}{2}\right)G\left(\frac{x_1-x_2}{2}\right)\right|^2.
\end{equation}

For performing the conditional coincidence measurements at the near-field plane we used 4x objective lenses to form the image of the center of the BBO-II onto the transverse plane of the fiber-couplers [see Fig.~(\ref{Fig2})(c)]. For doing this, we removed the slits used to scan the coincidence rate in the far-field plane and the lens $L_q$. The multi-mode fibers were replaced with single-mode fibers whose core diameter were $4.7\,\mu m$. The small size of the fibers core allows for post-selecting and measuring with high accuracy $\Delta^2(x_2|x_1)$ \cite{ExterNF2,Eisert}. Again, the effect of the transmittance function of the fiber over the broadening of the near-field distribution is negligible \cite{Kfact}. This can be easily checked through the calculation of the convolution between the transmittance function of the fiber with the predicted near-field distribution for our experimental configurations. By scanning transversely the single-mode fiber at mode-2, we recorded the coincidence conditional distribution at the near-field plane. An example of the results obtained is shown in Fig.~\ref{Fig3}(c) [Fig.~\ref{Fig3}(d)] for the second (sixth) state generated. The experimental results are represented by red points. The black dotted-line is the theoretical curve for these distributions arising from Eq.~(\ref{SPDC-pos}).

\begin{figure}[t]
\centering
\includegraphics[width=0.48\textwidth]{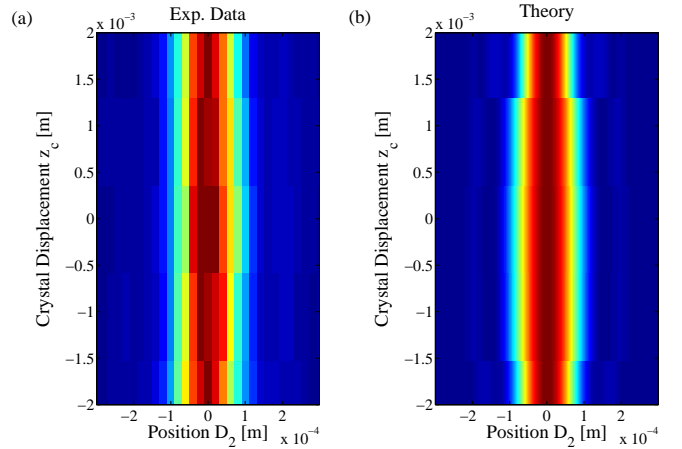}
\caption{(Color Online) (a) Experimental result and (b) theoretical prediction for the near-field conditional distribution while moving longitudinally the non-linear crystal (See the main text for details).\label{Fig4}}
\end{figure}

We have measured the coincidence conditional distribution of the down-converted photons by imaging the center of the non-linear crystal to the transverse plane of the detection system. However, as it has been shown in Ref. \cite{ExterNF2}, the conditional distribution at the near-field plane depends strongly of which part of the crystal is imaged at the detection system. This is specially relevant when thicker crystals are considered. In our case,  we have checked that the near-field distribution does not change significantly while considering different planes of our thin crystal to be imaged at the detection system. For doing this, we moved the crystal around the longitudinal central position $z_0=0$, imaging 5 different planes $z_c$ of the crystal at the detection plane. For each $z_c$, we recorded the conditional distribution at the near-field plane. The experimental results are shown in Fig.~(\ref{Fig4}). One can observe that our results are in agreement with the theoretical prediction, which takes into account our crystal length and our imaging system. A longitudinal crystal displacement around its center introduces a phase factor of $\exp[-i|\mathbf{q}_j^2|z/(2k)]$ onto Eq.~(\ref{SPDC-mom}). From our results, one can see that there is only a slight narrowing of the near-field conditional distribution such that this effect does not affect significantly our test of $W$.

\subsection{Testing $W$}

\begin{table*}[t]
\caption{Results of the conditional variances measured at the near- and far-field planes.\label{tabla2:resultados}}
\begin{ruledtabular}
\begin{tabular}{ccccc}
$P$ & $\Delta^2(x_2|0)_E\,\left[m^2\right]$ & $\Delta^2(q_2|0)_E\,\left[\frac{1}{m^2}\right]$ &$W_\mathrm{E}$ & $W_\mathrm{T}$\\ \hline
$0.1595$ & $(6.62\pm0.26)\times 10^{-10}$ & $(3.55\pm0.14)\times 10^{7}$ & $0.024\pm 0.002$ & $0.033$ \\
$0.3189$ & $(9.17\pm0.37)\times 10^{-10}$ & $(1.25\pm0.05)\times 10^{8}$ & $0.115\pm 0.009$ & $0.11$ \\
$0.4556$ & $(7.76\pm0.31)\times 10^{-10}$ & $(2.60\pm0.1)\times 10^{8}$ & $0.20\pm 0.02$ & $0.19$ \\
$0.7087$ & $(7.75\pm0.31)\times 10^{-10}$ & $(4.67\pm0.19)\times 10^{8}$ & $0.36\pm 0.03$ & $0.34$ \\
$0.7973$ & $(7.23\pm0.29)\times 10^{-10}$ & $(5.28\pm0.21)\times 10^{8}$ & $0.38\pm 0.03$ & $0.38$ \\
$0.9112$ & $(8.16\pm0.33)\times 10^{-10}$ & $(5.18\pm0.21)\times 10^{8}$ & $0.42\pm 0.03$ & $0.43$ \\
\end{tabular}
\end{ruledtabular}
\end{table*}

In order to test the $W$ witness, and then certify the non-gaussian feature of the CV spatial state of SPDC, we compute the variances of the conditional coincidence measurements at the near- and far-field planes for the six generated states. The results of all the conditional variances measured are shown in Tab.~\ref{tabla2:resultados}. The errors of the variances were obtained by minimizing the squared two-norm of the residuals between the analytical and experimental results (See Fig.~\ref{Fig3}).

\begin{figure}[ht]
\centering
\includegraphics[angle=0,width=0.4\textwidth]{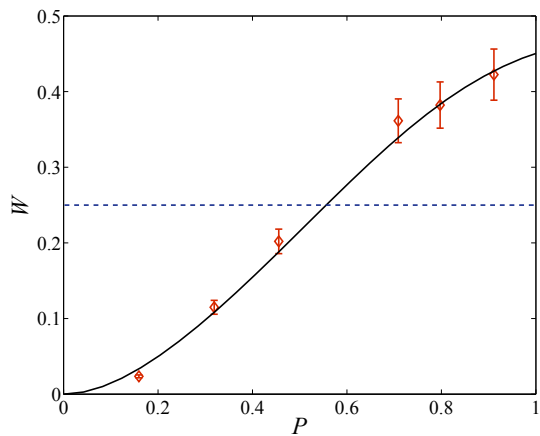}
\caption{(Color Online) Measurement of $W$ for the six generated states. The continuous (black) line corresponds to the theoretical value of the inequality for the spatial SPDC state. The dashed (blue) line shows the Gaussian bound. \label{Fig5}}
\end{figure}


Figure~\ref{Fig5} shows the values of the $W$ values for each state generated. The continuous (black) line indicate the predicted theoretical value of the inequality when we consider the spatial state of SPDC. The dotted (blue) line is the upper limit for this inequality for Gaussian states. For the first three entangled states, the values obtained for this inequality are below the Gaussian limit as predicted by theory. However, for the last three entangled states, there is a clear experimental violation of this bound. Due to momentum conservation, the spatial state of the SPDC process is always entangled \cite{Eberly} [See Fig.~\ref{Fig1}(a)] and, thus, we have a clear experimental demonstration of the non-Gaussianity of the CV two-photon spatial state of SPDC.

\section{Conclusion}\label{conc}
We have introduced a novel application for the EPR-steering inequality by showing that it can be used for witnessing the non-Gaussianity of CV quantum states. To demonstrated this we performed an experiment using the CV spatial state of entangled down-converted photons. Due to the phase-matching conditions of the SPDC process, the generated is state is naturally a pure entangled non-Gaussian state. A clear violation of the Gaussian bound of the EPR-steering inequality has been observed. Since non-Gaussian states are required for many new protocols of CV-QI, our work highlights the simplicity and relevance of using SPDC sources for new applications in CV quantum information processing.

\subsection*{Acknowledgments}
We thank C. H. Monken for discussions of this paper. This work was supported by Grants FONDECYT 1120067, Milenio P10-030-F and CONICYT FB0824/2008. E. S. G. and G. C. acknowledge the financial support of CONICYT. W. A. T. N.  thanks CNPq (Brazil) for financial support.


\end{document}